\documentclass[a4paper,11pt]{article}
\pdfoutput=1 

\usepackage{jcappub} 

\usepackage[T1]{fontenc} 
\usepackage{graphicx}
\usepackage[titletoc,title]{appendix}
\usepackage{bm}

\title{Fast Generation of Covariance Matrices for Weak Lensing}

\author[a]{R. J. Sgier,}
\author[a]{A. Réfrégier,}
\author[a]{A. Amara}
\author[a]{and A. Nicola}

\affiliation[a]{Institute for Particle Physics and Astrophysics, Department of Physics, ETH Zurich,\\Wolfgang-Pauli-Strasse 27, 8093 Zurich, Switzerland}

\emailAdd{raphael.sgier@phys.ethz.ch}
\emailAdd{alexandre.refregier@phys.ethz.ch}
\emailAdd{adam.amara@phys.ethz.ch}
\emailAdd{andrina.nicola@phys.ethz.ch}

\abstract{
Upcoming weak lensing surveys will probe large fractions of the sky with unprecedented accuracy. To infer cosmological constraints, a large ensemble of survey simulations are required to accurately model cosmological observables and their covariances. 
We develop a parallelized multi-lens-plane pipeline called \textsc{UFalcon}, designed to generate full-sky weak lensing maps from lightcones within a minimal runtime. It makes use of L-PICOLA (Howlett \textit{et al.} \cite{PICOLA}), an approximate numerical code, which provides a fast and accurate alternative to cosmological $N$-Body simulations. The \textsc{UFalcon} maps are constructed by nesting 2 simulations with mass resolution of about $5 \, \times \, 10^{12}$ and $2\, \times \, 10^{13}$  $h^{-1} \, \text{M}_\odot$ covering a redshift-range from $z=0.1$ to $1.5$ without replicating the simulation volume. We compute the convergence and projected overdensity maps for L-PICOLA in the lightcone or snapshot mode. The generation of such a map, including the L-PICOLA simulation, takes about 3 hours walltime on 220 cores. We use the maps to calculate the spherical harmonic power spectra, which we compare to theoretical predictions and to \textsc{UFalcon} results generated using the full $N$-Body code GADGET-2. We then compute the covariance matrix of the full-sky spherical harmonic power spectra using 150  \textsc{UFalcon} maps based on L-PICOLA in lightcone mode. We consider the PDF, the higher-order moments and the variance of the smoothed field variance to quantify the accuracy of the covariance matrix, which we find to be a few percent for scales $\ell \sim 10^2$ to $10^3$. We test the impact of this level of accuracy on cosmological constraints using an optimistic survey configuration, and find that the final results are robust to this level of uncertainty. The speed and accuracy of our developed pipeline provides a basis to also include further important features such as masking, varying noise and will allow us to compute covariance matrices for models beyond $\Lambda$CDM.}
 
\begin{document}
\maketitle
\flushbottom

\section{Introduction}
\label{intro}

The standard model of cosmology predicts the formation of large-scale structures through gravitational amplification of density perturbations. It describes our universe as a spatially flat $\Lambda$CDM cosmology, which is mostly filled with dark energy and dark matter. The advent of modern cosmological surveys such as the cosmic microwave background (CMB) measurement (e.g. Planck Collaboration \cite{Planck}) and large-scale structure surveys established the standard model of cosmology. Understanding the nature of the model and its main constituents, namely dark energy and dark matter, is thus one of the greatest challenges in physics today. Numerous surveys pursue this goal by probing large cosmic volumes and with increased sensitivity, such as the Dark Energy Survey (DES\footnote{http://www.darkenergysurvey.org}), the Dark Energy Spectroscopic Instrument (DESI\footnote{http://desi.lbl.gov}), the Large Synoptic Survey Telescope (LSST\footnote{http://lsst.org}), Euclid\footnote{http://sci.esa.int/euclid/}, the Wide Field Infrared Survey Telescope (WFIRST\footnote{http://wfirst.gsfc.nasa.gov}) and the Kilo-Degree Survey (KiDS\footnote{http://kids.strw.leidenuniv.nl/}). In particular, the results of the DES science verification data (DES SV)~\cite{Jarvis, Becker, Abbott} and from the first year survey (DES Y1)~\cite{Y11, Y12, Y13, Y14, Y15, Y16, Y17, Y18, Y19, Y110} just mark the beginning of the era of precision large-scale structure surveys.

The effort is based on analyzing different cosmological probes such as weak gravitational lensing, which probes the matter structure between the observer and background light sources that act as lenses and thus alter photons geodesics (e.g. Bartelmann \& Schneider~\cite{Bartelmann2}; Kilbinger~\cite{Kilbinger}; Bartelmann \& Maturi~\cite{Bartelmann1} for a review). The measured signal consists of deformed shapes of distant galaxies, which allows us to infer the density fluctuations in the foreground matter distribution. The distance to the sources and the growth rate of the density fluctuations also gives information about dark energy. Understanding the properties of structure formation and accelerated expansion gives thus direct information about the nature of dark matter and dark energy, respectively.

The amount of information retrievable from large-scale structure surveys scales as the cube of the largest wavenumber reliably observable. Given the high accuracy of present and future weak lensing surveys, it is thus of increasing importance to have an accurate analytical prediction of large-scale structure formation at the smallest scales possible in order to correctly interpret the data and to obtain unbiased constraints on cosmological parameters. While numerous analytical attempts at using high-order perturbation theory (e.g. Bernardeau \textit{et al.} \cite{Bernardeau} for a review) give accurate results in the mildly-nonlinear regime, they are typically not adapted to describe the small scales resolved by modern surveys. The current lack of high-precision analytical methods to accurately describe the nonlinear regime of structure formation thus makes high-resolution cosmological $N$-Body simulations the main tool at hand for this purpose. Such numerical simulations need to have a high resolution to correctly describe scales probed by surveys (e.g. the MICE Grand Challenge simulation \cite{MICE, MICE1, MICE2}; the blind cosmology challenge (BCC) \cite{risa}). 

High accuracy is not only necessary for the power spectrum but also for its covariance (Takahashi \textit{et al.} \cite{Takahashi1}). The covariance matrix encodes the statistical uncertainty of the power spectrum, which for the case of Gaussian density fluctuations has only diagonal elements for the case of a full-sky survey. On small scales, the covariance matrix develops non-vanishing off-diagonal elements due to mode coupling in the non-linear regime of structure formation, which is described by the correlation between different modes (e.g. Scoccimarro \textit{et al.} \cite{Scoccimarro}; Meiksin \& White \cite{White}; Smith \cite{Smith}). Non-Gaussian contributions to the covariance matrix are known to affect the precision of cosmological parameter estimation and therefore need to be taken into account in large-scale structure surveys (Cooray \& Hu \cite{Cooray}). In order to obtain a well-converged estimate of the covariance matrix, a large sample of simulations is needed (Dodelson \& Schneider \cite{Dodelson}; Taylor \textit{et al.} \cite{Taylor}; Percival  \textit{et al.} \cite{Percival}). Since the generation of such a large sample of mocks is limited by the computational power available, various faster alternative methods to fully realized $N$-Body codes have been developed. Analytical methods based on the halo model have been developed und used to generate mock catalogues for DES (Eifler \textit{et al.} \cite{Eifler}; Krause \& Eifler \cite{Krause}). Further alternative methods include PINOCCHIO (Taffoni \textit{et al.} \cite{Taffoni}), Quick Particle Mesh Simulations (White \textit{et al.} \cite{White2}) or the Comoving Lagrangian Acceleration method (COLA; Tassev \textit{et al.} \cite{COLA}; Koda \textit{et al.} \cite{COLA1}). In fact, it has been shown in Howlett \textit{et al.} \cite{PICOLA} that the approximate code L-PICOLA based on the COLA method is able to reproduce the matter power spectrum generated using the full $N$-Body code GADGET-2 (Springel \textit{et al.} \cite{Springel1, Springel2}) in a much shorter runtime.

Cosmological probes such as weak gravitational lensing led to the establishment of past-lightcone algorithms applied to $N$-Body simulations (e.g. Teyssier \textit{et al.} \cite{Teyssier}; Vale \textit{et al.} \cite{Vale}; Hilbert \textit{et al.} \cite{Hilbert}; Takahashi \textit{et al.} \cite{Takahashi}). The obtained weak lensing maps are then analyzed by computing statistical quantities such as the spherical harmonic power spectrum of the convergence (e.g. Sato \textit{et al.} \cite{sato}; Kiessling \textit{et al.} \cite{Kiessling}; Harnois-Déraps \textit{et al.} \cite{Harnois1}; Patton \textit{et al.} \cite{Patton}; Giocoli \textit{et al.} \cite{Giocoli}; Izard \textit{et al.} \cite{COLA3}), the 1-point probability distribution (e.g. Takahashi \textit{et al.} \cite{Takahashi2}; Patton \textit{et al.} \cite{Patton}) and higher-order statistics.

The goal of this paper is to introduce a fast past-lightcone pipeline called \textsc{UFalcon} (Ultra Fast Lightcone), which produces full-sky weak lensing maps based on the approximate code L-PICOLA. Our pipeline is primarily designed to resolve angular scales up to $\ell \sim 10^3$ for wide field surveys such as DES \cite{Jarvis, Becker, Abbott, Y11, Y12, Y13, Y14, Y15, Y16, Y17, Y18, Y19, Y110} and cover large areas (e.g. 5000 $\mathrm{deg}^2$ for DES) and has a runtime of $\sim 3$ hours walltime for one realization of the matter density field and its corresponding full-sky weak lensing map.

A discussion of the $N$-Body simulations used and our \textsc{UFalcon}-pipeline is presented in section \ref{nummeth}. In section \ref{weaklensingtheory} we review the derivation of the convergence mass and overdensity map, which are the two weak lensing maps used for our statistical analysis. We perform a statistical analysis in section \ref{results} of our results by computing the spherical harmonic power spectrum, the covariance matrix, the probability distribution function and higher-order statistics. The power spectra and covariance matrix are then used to constrain the cosmological parameters in section \ref{mcmc}. A discussion of our analysis is given in section \ref{conclusion}.
\\

\section{Numerical Methods}
\label{nummeth}
The gravitational collapse of matter on small scales $(\lesssim 10 \, \mathrm{Mpc})$ and late times is a highly nonlinear process. Linear perturbation theory is thus insufficient at describing this complex process. Even predictions from modern higher-order perturbation theories and approaches based on a halo-model have limited accuracy in describing nonlinear scales. Cosmological $N$-Body simulations are thus currently the most accurate method available to describe different dynamical ranges of structure formation, which is crucial to correctly interpret the data obtained by modern surveys. Furthermore, a large number of realizations are needed to reduce the statistical error in the estimation of the covariance of weak lensing observables. Thus, depending on the desired resolution of the simulation and number of realizations, the limiting factor usually is the available computing power. 
\\
\subsection{$N$-Body Simulations}
Cosmological $N$-Body simulations come in numerous variations. Particle-Mesh (PM) codes (e.g. Klypin \textit{et al.} \cite{Klypin1, Klypin2}; Hockney \textit{et al.} \cite{Hockney}) rely on solving the Poisson equation in Fourier space to easily obtain the gravitational potential and force (Bagla \textit{et al.} \cite{Bagla}). The use of only one mesh in the simulation for both the density and the potential lowers the number of operations per particle per timestep for long-range force calculations. More complex simulations such as TREE-PM codes (e.g. Bode \textit{et al.} \cite{Ostriker}; Appel \textit{et al.} \cite{Appel}; Barnes \textit{et al.} \cite{Barnes_Hut}; Hernquist \textit{et al.} \cite{Hernquist}) combine the PM approach with the more accurate TREE code: In this way the simulation volume is divided into cells, subcells and particles. Each cell has its total mass and center of mass, which is used for the force calculation if the cell is sufficiently far away and can be treated as a single entity (Bagla \textit{et al.} \cite{Bagla}).

Further approximate simulations for solving for large-scale structure are known as COLA methods (Tassev \textit{et al.} \cite{COLA}; Koda  \textit{et al.} \cite{COLA1}), which are based on a modified leapfrog point-mesh algorithm using first- and second-order Lagrangian Perturbation Theory (LPT) and perform the calculations in a frame comoving with the trajectories. In this framework, the linear growth factor is directly computed for large scales, while the dynamics on small scales are computed using a PM code. The $N$-Body code COLA allows us to trade accuracy on small scales by using only a few timesteps to gain computational speed, while using the exact second-order LPT result on large scales.

Recent developments gave rise to the $N$-Body code L-PICOLA (lightcone-enabled parallel integration COLA; Howlett \textit{et al.} \cite{PICOLA}) and first applications thereof (e.g. Izard \textit{et al.} \cite{COLA2, COLA3}; Patton \textit{et al.} \cite{Patton}). We use the L-PICOLA code for the present work, which has been shown to be several orders of magnitude faster than conventional $N$-Body codes such as GADGET-2.
\\
\subsection{Simulation Setup}

We compare the results from applying our new gravitational lensing lightcone pipeline using L-PICOLA to the ones using the $N$-Body code GADGET-2. This starts by choosing $\Lambda$CDM cosmological parameters $h = 0.7$, $\Omega_m = 0.276$, $\Omega_b = 0.045$, $n_s = 0.961$ and $\sigma_8 = 0.811$ to generate the initial conditions at an initial redshift $z_\text{init}=9$ for both simulation suites. This specific choice for the initial redshift works particularly well for COLA and has been suggested by Howlett \textit{et al.} \cite{PICOLA}. We use the modified COLA timestepping for kick and drift with the value $nLPT= -2.5$ as suggested by Tassev \textit{et al.} \cite{COLA}. The initial conditions for L-PICOLA are generated by the code itself before the  gravitational evolution of the dark matter particles begin, whereas we use the publicly available code MUSIC (MUlti-Scale Initial Conditions; Hahn \textit{et al.} \cite{HahnAbel}) to generate the initial conditions used for GADGET-2. Both simulations are set up using an Eisenstein \& Hu \cite{EisensteinHu} transfer function. We run 150 L-PICOLA simulations in the lightcone mode for each of the two simulation setups given in Table \ref{table2}. This should allow us to reach the precision regime of $1-10\%$ and the runtimes are comparable to what is needed for roughly four comparable GADGET-2 runs. We construct the full-sky past-lightcone by nesting the two simulation volumes (see section \ref{pastlightcone_construction}). Furthermore, we also produce one GADGET-2 and one L-PICOLA realization in the usual snapshot output mode for the two simulation setups, where the positions and velocities of all the particles in the simulation volume are stored at a desired redshift. Here we choose the simulation to output snapshots at a redshift spacing of $\delta z = 0.02$.
\\
\begin{table}
\centering
\begin{tabular}{cccccc}
\hline
$L_\text{Box}$ & $N_\text{Part}$  & $N_\text{Mesh}$ & mass resolution & $z$-range & $z_\text{init}$ \\
$(h^{-1} \, \text{Mpc})$ & & & $(h^{-1} \, \text{M}_\odot)$  & &  \\
\hline
4200 & 1024 & 2048 & $5.2 \, \times \, 10^{12}$ & 0.1 - 0.8 & 9 \\
6300 & 1024 & 2048 & $1.8 \, \times \, 10^{13}$ & 0.8 - 1.5 & 9 \\
\hline
\end{tabular}
\caption{Parameters used to simulate the matter density field.}
\label{table2}
\end{table}

\subsection{Convergence and Overdensity Maps}
\label{weaklensingtheory}

The following treatment of the convergence and galaxy overdensity mass map in the Born approximation is based on the work done in Teyssier \textit{et al.} \cite{Teyssier}, Pires \textit{et al.}~\cite{Pires} and Schmelzle \textit{et al.}~\cite{Jorit}. The overdensity can be related to the projected convergence field for a single source redshift located at $z_s$ through
\begin{equation}
\kappa (\hat{n}) = \frac{3}{2} \Omega_\text{m} \int_0^{z_s} \frac{\text{d}z}{E(z)} \frac{\mathcal{D}(z)\mathcal{D}(z, z_s)}{\mathcal{D}(z_s)} \frac{1}{a(z)} \delta \left(\frac{c}{H_0}\mathcal{D}(z)\hat{n}, z \right)\, ,
\end{equation}
where the dimensionless comoving (and radial) distance is given by $\mathcal{D}(z) = (H_0 / c) \chi (z)$. Invoking the Born approximation and the fact that one has to compute the convergence from the matter field coming from simulations at discrete redshifts, one can approximate the above equation as a discrete sum over redshift slices
\begin{equation}
\label{kappa_1}
\kappa (\theta_\text{pix}) \approx \frac{3}{2} \Omega_\text{m} \sum_b W_b \frac{H_0}{c} \int_{\Delta z_b} \frac{c\, \text{d}z}{H_0 E(z)}  \delta \left(\frac{c}{H_0}\mathcal{D}(z)\hat{n}_\text{pix}, z \right)\, ,
\end{equation}
where $\theta_\text{pix}$ is the position on the pixelized sphere and the slice-related weights of the convergence $W_b$ contain the source redshifts. In the case of a single delta-distributed source redshift at $z_s$, the weight can be written as
\begin{equation}
W_b^{\text{delta}} = \left( \int_{\Delta z_b} \frac{\text{d}z}{E(z)} \frac{\mathcal{D}(z) \mathcal{D}(z, z_s)}{\mathcal{D}(z_s)} \frac{1}{a(z)} \right) / \left( \int_{\Delta z_b} \frac{\text{d}z}{E(z)} \right) \, .
\end{equation}
The integral over the density contrast can be recast as an integral over comoving distance
\begin{equation}
\int_{\Delta \chi_b} \text{d}\chi  \delta \left(\chi \hat{n}_\text{pix}, \chi \right) = \int_{\Delta \chi_b} \text{d}\chi \frac{\rho\left(\chi \hat{n}_\text{pix}, \chi \right)}{\bar{\rho}} - \Delta \chi_b \, .
\end{equation}
Integrating out the length of one shell $\Delta \chi_b$ along the line of sight in the density gives an expression for the surface density
\begin{equation}
\int_{\Delta \chi_b} \text{d}\chi  \rho \left(\chi \hat{n}_\text{pix}, \chi \right) = \sigma \left(\chi \hat{n}_\text{pix}, \chi \right) = \frac{m_p \cdot n_p  \left(\text{pix}, \Delta \chi_b \right)}{l^{2}_{\chi_b}} \, ,
\end{equation}
where, in the last step, we used the fact that we are considering discrete particles on a pixelized sphere: The mass of a particle is given by $m_p$, whereas $n_p$ represents the number of particles in a specific pixel on the shell $\Delta \chi_b$. The area of a pixel on the sphere with radius $\chi_b$ is given by $l^{2}_{\chi_b} = 4 \pi \chi_b^2 / N_\text{pix}$, where $N_\text{pix}$ is the number of pixels. Furthermore, the average density can be written as the total mass of all particles in the simulation divided by the simulation volume $\bar{\rho} \approx m_p \cdot N_\text{part}^\text{sim} / V_\text{sim}$. The convergence map given by equation (\ref{kappa_1}) can now be written as
\begin{equation}
\kappa (\theta_\text{pix}) \approx \frac{3}{2} \Omega_m \sum_b W_b \frac{H_0}{c} \left[ \frac{N_\text{pix}}{4 \pi} \frac{V_\text{sim}}{N_\text{part}^\text{sim}} \left( \frac{H_0}{c} \right)^2 \frac{n_p (\theta_\text{pix}, \Delta \chi_b)}{\mathcal{D}^2 (z_b)}  - \left( \frac{c}{H_0} \Delta \mathcal{D}_b \right)\right] \, ,
\end{equation}
and the overdensity is given by
\begin{equation}
\delta (\theta_\text{pix}) \approx \sum_b \left[ \frac{N_\text{pix}}{4 \pi} \frac{V_\text{sim}}{N_\text{part}^\text{sim}} \left( \frac{H_0}{c} \right)^2 \frac{n_p (\theta_\text{pix}, \Delta \chi_b)}{\mathcal{D}^2 (z_b)}  - \left( \frac{c}{H_0} \Delta \mathcal{D}_b \right) \right] / \sum_b \left( \frac{c}{H_0}  \Delta \mathcal{D}_b \right) \, ,
\end{equation}
where $\Delta \mathcal{D}_b$ is the thickness of a redshift-slice in dimensionless comoving coordinates.
\\
\subsection{Past-Lightcone Construction}
\label{pastlightcone_construction}

The computation of the full-sky convergence map is based on the past-lightcone, which represents the lensed photons traveling from the source at redshift $z_s$ to the observer located at $z=0$, i.e. the photon geodesics through large-scale structure. Therefore, the construction of the lightcone needs to take into account the matter field at different past times acting as gravitational lenses for the photons. In order to build an all-sky past-lightcone from $N$-body simulation outputs, we begin by fixing the observer at the center of the simulation volume at $z=0$. This allows us to construct a lightcone up to our desired source redshift, which we choose to be the edge of the simulation volume. In this way, we avoid the construction of the lightcone by replicating smaller boxes, which would result in sampling the same modes multiple times. Depending on the simulation configuration, such replication effects could have an impact on the spherical harmonic power spectrum and covariance matrix, even when the replicated boxes are randomized in order to break the artificial correlation (e.g. Howlett \textit{et al.} \cite{PICOLA}). Furthermore, using large enough boxes ensures that super-survey modes are correctly captured, i.e. that no issues concerning super-sample covariance occur (Takada \textit{et al.} \cite{supersample}).

The volume between the observer and the source is sliced without gaps into comoving concentric spherical shells of thickness $\Delta \chi_b = \chi(z_b + \Delta z) - \chi (z_b)$, which contain the particles with radial coordinates

\begin{equation}
\chi(z_b)  \leq r_p \leq \chi(z_b + \Delta z) \quad ,
\end{equation} 
where we choose a redshift-shell thickness of $\Delta z = 0.01$ and all the particles in the snapshot $b$ correspond to redshift $z_b$. Running L-PICOLA in the lightcone-mode generates one output consisting of the past-lightcone of the observer, i.e. a spherical arrangement of the particles at desired output-redshifts.

\begin{figure}[htbp!]
\centering
\includegraphics[width=10cm]{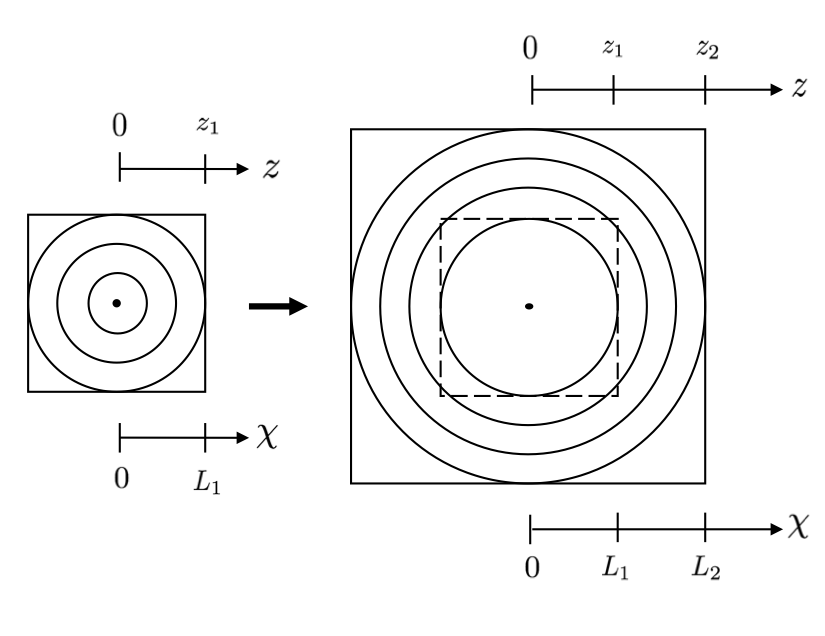}
\caption{Left: As a first step the lightcone is constructed by using the smaller box with side-length $L_1$. Right: Once the edge of the smaller box is reached, the construction is continued in a bigger box with side-length $L_2\, (> L_1)$. The observer is located at the center of the boxes.} \label{sketch}
\end{figure}

In our current implementation, we construct an all-sky past-lightcone from us up to a single source located at redshift $z_s = 1.5$ or for sources with a specific redshift distribution $n(z)$. The high $z$ source imposes the need for a large simulation volume, which suggests the use of a large number of particles and mesh-size in order to reach a desired resolution of $\ell \sim 10^3$. This would increase the needed computational requirements for a large number of realizations, which is difficult in practice. Instead, a nesting scheme (e.g. White \& Hu \textit{et al.} \cite{WhiteHu}; Busha \textit{et al.} \cite{risa}) is adopted to maintain acceptable computational costs and high resolution up to certain source redshifts without having to replicate the volume: Smaller boxes are placed within larger boxes, while maintaining the same number of particles and mesh-size. When the lightcone, starting at the observer, reaches the edge of the smallest box, it continues by using concentric shells in the next larger box.  A sketch of the used nesting scheme is shown in Figure \ref{sketch}.

\begin{figure}[htbp!]
\centering
\includegraphics[width=15cm]{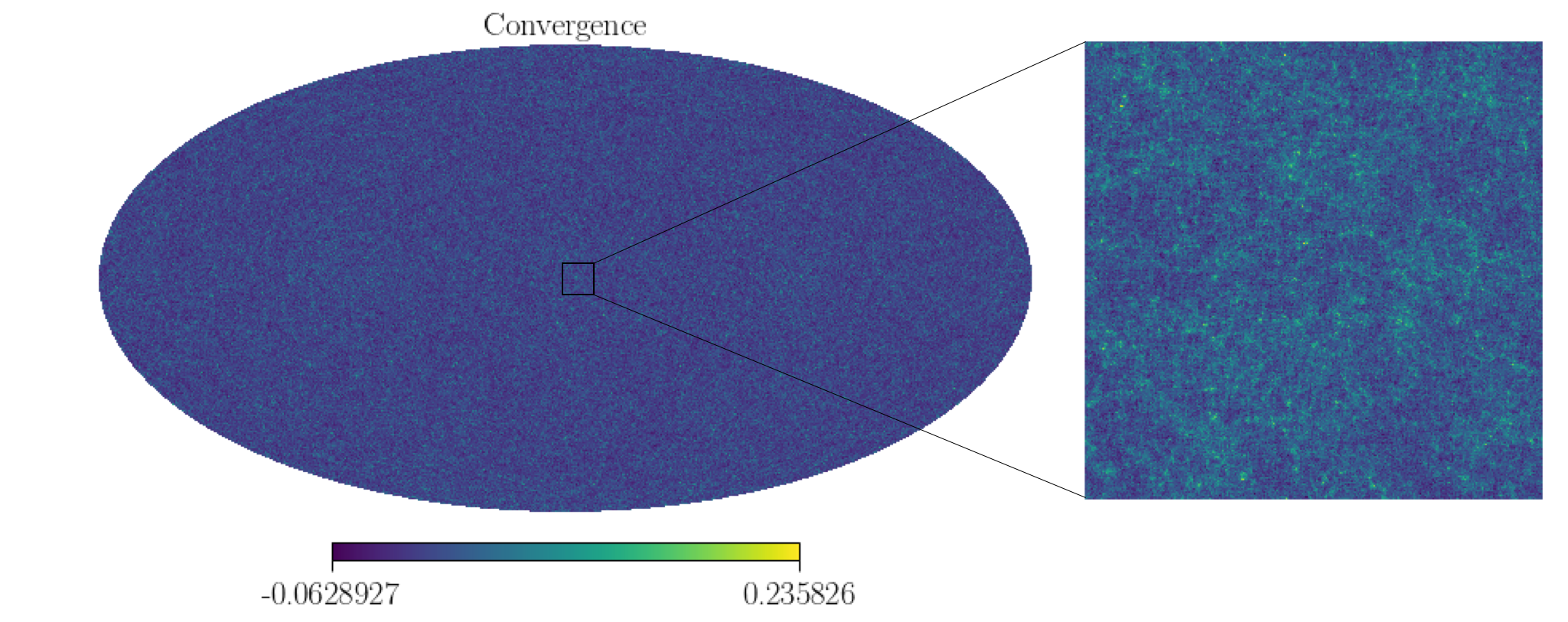}
\caption{Full-sky convergence map computed using the \textsc{UFalcon}-pipeline by nesting two L-PICOLA simulations with box-sizes $L_1 = 4200\, h^{-1}\,\text{Mpc}$ and $L_2 = 6300\, h^{-1}\,\text{Mpc}$ (see Table \ref{table2}) covering the redshift range $z = 0.1 - 1.5$. The zoom-in quadrant shows the filamentary, nonlinear structure on smaller scales.} \label{kappa_zoom}
\end{figure}

\newpage
In order to find a trade-off between the needed accuracy and computational cost, we construct the lightcone by nesting the simulations with two box-sizes $L_1 = 4200\, h^{-1}\,\text{Mpc}$ and $L_2 = 6300\, h^{-1}\,\text{Mpc}$ (see Table \ref{table2}), which allows us to cover a redshift range from $z = 0.1$ to $z_s = 1.5$. This range is sliced into shells of thickness $\Delta z = 0.01$, which is thin enough for the Born approximation to hold. We find that the use of thinner redshift-slices do not provide more accuracy as far as the spherical harmonic power spectrum is concerned. Both the runtime for simulating the density field ($\sim 2$ hours for a mass resolution of $\sim 10^{13} \, h^{-1} \, \text{M}_\odot$ using 220 cores) and for the past-lightcone construction ($\sim 1$ hour) are greatly enhanced through parallelization on high-performance clusters. The parallelization of the lightcone construction is achieved by dividing the obtained simulation volume into subvolumes, which can be processed individually to construct the past-lightcone. Since the subvolumes contain less particles than the full simulation volume, the construction is computationally less expensive. The lightcone is thus computed for each subvolume simultaneously and the obtained maps are then combined together. The pixelization procedure on the sphere is done using \textsc{HEALPix}\footnote{http://healpix.sourceforge.net}, which returns $12\cdot \text{\tt NSIDE}^2$ pixels for a given resolution {\tt NSIDE}. Figure \ref{kappa_zoom} shows an example of a full-sky map of the convergence using our \textsc{UFalcon}-pipeline described above.

\section{Statistical Analysis}
\label{results}
\subsection{Power Spectrum}
\label{powerspectra}
\begin{figure}[htbp!]
\centering
\includegraphics[width=10.0cm]{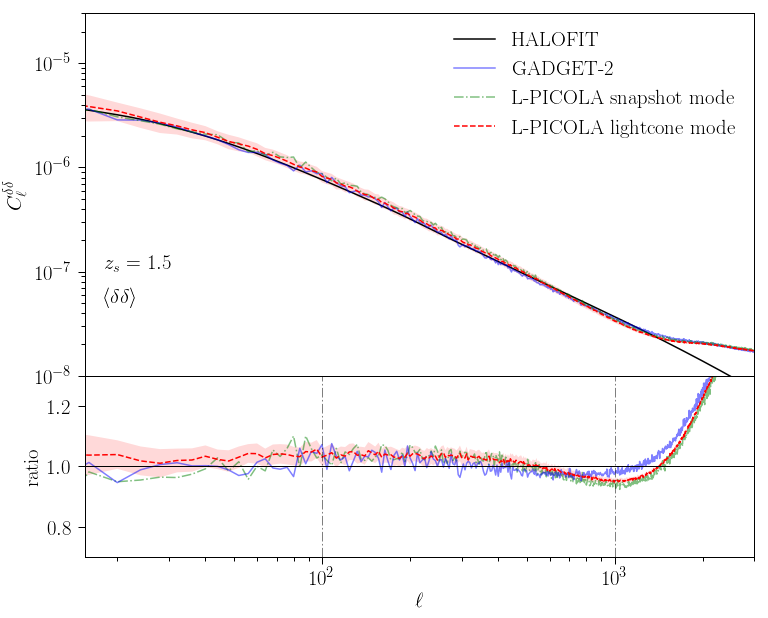}
\caption{Full-Sky spherical harmonic power spectrum for the convergence (upper panel) and the galaxy overdensity (lower panel) with a $\Delta \ell = 4$ binning. The mean of 150 L-PICOLA realizations in the lightcone mode (red dashed) is compared to one L-PICOLA snapshot mode run (green dot-dashed) and one GADGET-2 run (blue solid). All three power spectra from simulations are divided by the nonlinear HALOFIT prediction (black solid) in the lower subpanels. The red area represents the $1\sigma$-error of the 150 L-PICOLA lightcone mode simulations.} \label{PS_delta_kappa}
\end{figure}

We first study the spherical harmonic power spectrum defined as
\begin{equation}
\left< q_{\ell m} q^{*}_{\ell' m'} \right> = \delta_{\ell \ell'} \delta_{m m'} C_\ell ^{q} \, ,
\end{equation}
where we choose the fields $q = \kappa$ or $\delta$ and $q_{\ell m}$ is the spherical harmonic coefficient. The multipole moment $\ell$ corresponds to the inverse of the angular scale and the integer $m$ lies in the range $-\ell < m < \ell$. The brackets $\left< ... \right>$ represent an ensemble average. The harmonic analysis on the sphere is performed using \textsc{ANAFAST}, which discretizes the spherical harmonics up to $\ell_\text{max} = 3 \cdot \text{\tt NSIDE} - 1$. Using a resolution of {\tt NSIDE}=1024 returns us modes up to $\ell < 3071$, which is sufficient for our purposes. We compare the full-sky convergence spherical harmonic power spectrum from L-PICOLA computed in the lightcone mode (mean of 150 realizations) and snapshot mode with those from high-resolution GADGET-2 simulations (in snapshot mode) using the same relevant simulation parameters (see Table \ref{table2}), initial conditions, seed and nesting-scheme. The results for a single source redshift at $z_s = 1.5$ are plotted in Figure \ref{PS_delta_kappa} for the convergence and the overdensity and a multipole binning of $\Delta \ell = 4$ together with the revised nonlinear HALOFIT power spectrum predictions (Takahashi \textit{et al.} \cite{Halofit}; Smith \textit{et al.} \cite{smith}) computed within \textsc{PyCosmo} (Refregier \textit{et al.} \cite{pycosmo}) based on the Limber approximation.

Our result drawn from the L-PICOLA lightcone simulations $C_{\text{lightcone}}^{\kappa \kappa}$ agrees with the theoretical prediction to within $5\%$ on scales $10^2 < \ell < 10^3$. The power spectrum computed for the overdensity $C_{\text{lightcone}}^{\delta \delta}$ also agrees to within $5\%$ of the theory prediction for multipoles $10^2 < \ell < 1600$. We see that the L-PICOLA power spectra slightly exceed the theory prediction and the GADGET-2 power spectra for scales $10^2 < \ell < 700$, whereas they underestimate the power on smaller scales. The steep divergence of all power spectra computed from $N$-Body simulations around $\ell \sim 1200$ for $C_{\ell}^{\kappa \kappa}$ and around $\ell \sim 10^3$ for $C_{\ell}^{\delta \delta}$ is due to the shot noise arising in the simulation at the mass resolution we are working at.

In order to exclude effects coming from our past-lightcone construction in the comparison, i.e. effects arising due to the Born-approximation, we also compare the spherical harmonic power spectrum from the L-PICOLA simulation to the one from GADGET-2. We find that the results for the convergence and overdensity power spectra based on the L-PICOLA lightcone mode agree within 5\% for $10^2 < \ell < 10^3$ to the result drawn from the GADGET-2 simulation.

\subsection{Covariance Matrix}
\label{cov_matrix}

\begin{figure}[htbp]
  \begin{minipage}[b]{0.47\linewidth}
    \centering
    \includegraphics[width=\linewidth]{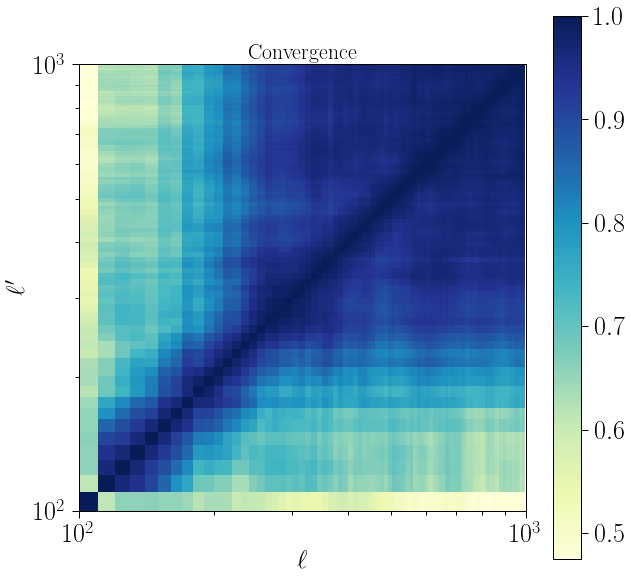}
  \end{minipage}
  \hspace{0.5cm}
  \begin{minipage}[b]{0.47\linewidth}
    \centering
    \includegraphics[width=\linewidth]{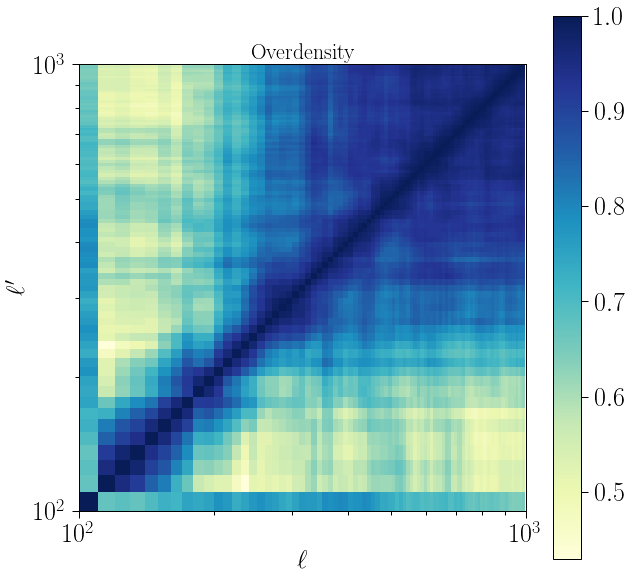}
  \end{minipage}
  \caption{Correlation coefficient matrix of the power spectrum $C_\ell$ for angular scales $10^2 < \ell < 10^3$ calculated from 150 L-PICOLA realizations for $z_s = 1.5$ and a binning of $\Delta \ell = 10$. The left plot corresponds to the convergence, while the right plot corresponds to the overdensity.}
  \label{corr_matrix}
\end{figure}

We compute the covariance of the power spectra using 150 full-sky L-PICOLA simulations, which can be written as (e.g. Nicola \textit{et al.} \cite{Andrina1})
\begin{equation}
\text{cov} ( \ell, \ell' ) = \left< C_\ell^{q} - \left< C_\ell^{q} \right> \right> \left< C_{\ell'}^{q} - \left< C_{\ell'}^{q} \right> \right> \, ,
\end{equation}
where $\left< ... \right>$ represents the average over all 150 L-PICOLA realizations in the lightcone mode. In order to better quantify the correlation between different multipoles it is useful to use the correlation coefficient defined as (Blot \textit{et al.} \cite{Blot})
\begin{equation}§
\text{corr} (\ell, \ell') = \frac{\text{cov} (\ell, \ell')}{\sqrt{\text{cov} (\ell, \ell) \text{cov} (\ell', \ell')}} \, ,
\end{equation}
such that the correlation between different modes ranges between $\text{corr} = +1$ (maximum correlation) and $\text{corr} = -1$ (maximum anti-correlation). Figure \ref{corr_matrix} shows the correlation coefficient matrices for the power spectum $C_\ell$ calculated using the L-PICOLA lightcone mode for a multipole binning of $\Delta \ell = 10$.
The mode coupling caused by the non-linearities at small scales and late times is captured by the off-diagonal elements of the covariance matrix. As visible in the figures, the correlation of different multipoles is small for multipoles smaller than $\ell \sim 2 \,\times \, 10^2$. However, the convergence shows a larger correlation than the overdensity in this regime. For scales larger than $\ell \sim 2.5 \,\times \, 10^2$, stronger correlations of different angular modes appear.

Note that in the case of a realistic survey setup one could use one full-sky map multiple times to obtain several realizations of the survey area. The use of 150 full-sky maps would thus be enough to obtain a well-converged estimate of the covariance matrix depending on the survey area.

\begin{figure}[htbp!]
\centering
\includegraphics[width=10.0cm]{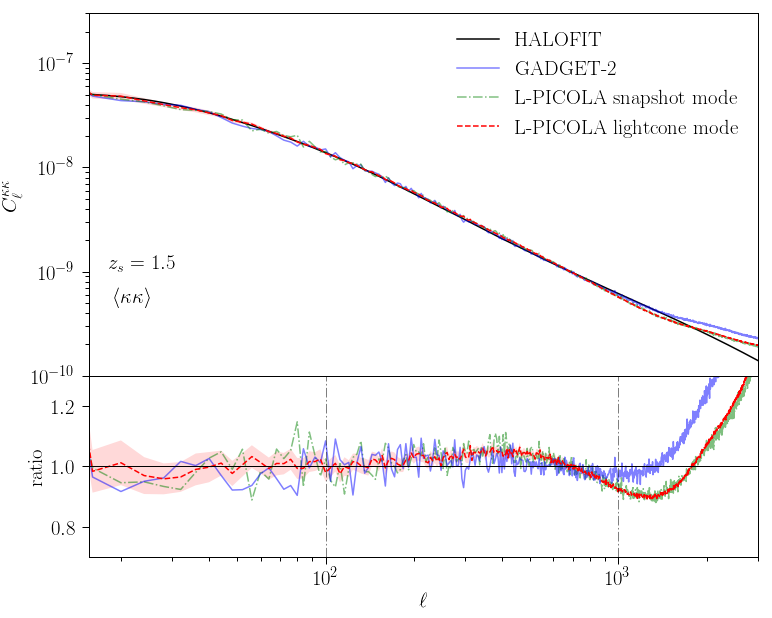}
\end{figure}

\subsection{Probability Distribution Function}
\begin{figure}[htbp!]
\centering
\includegraphics[width=12cm]{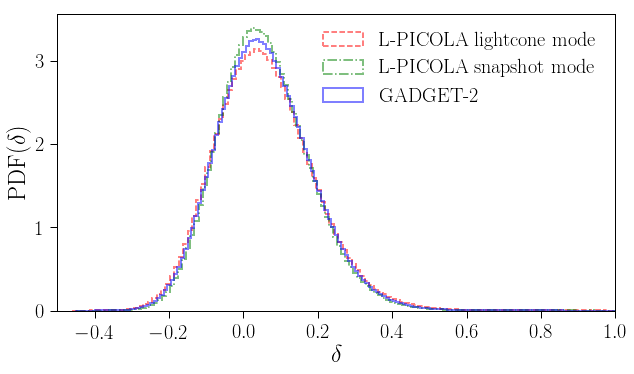}
\caption{Normalized convergence (upper plot) and overdensity (lower plot) PDF using L-PICOLA lightcone mode (red dashed) and snapshot mode (green dot-dashed) compared to using GADGET-2 (blue solid). All three maps have been smoothed using a Gaussian beam with an r.m.s. of 10.8 arcmin, which corresponds to $\ell \sim 10^3$.} \label{PDF_delta_kappa}
\end{figure}

To assess the accuracy of our covariance matrix estimate, we compare the probability distribution function (PDF, i.e. 1-point distribution) of the constructed maps from L-PICOLA to the ones from GADGET-2. We analyze our computed weak lensing maps on angular scales of our interest $10^2 < \ell < 10^3$. Figure \ref{PDF_delta_kappa} shows the PDF of the convergence and overdensity maps after smoothing using a Gaussian beam with an r.m.s. radius of $10.8$ arcmin corresponding to multipole $\ell \sim 10^3$ on the sphere. We find that the maps generated using L-PICOLA and GADGET-2 smoothed on this scale show very similar distributions.

\begin{figure}[htbp!]
\centering
\includegraphics[width=12cm]{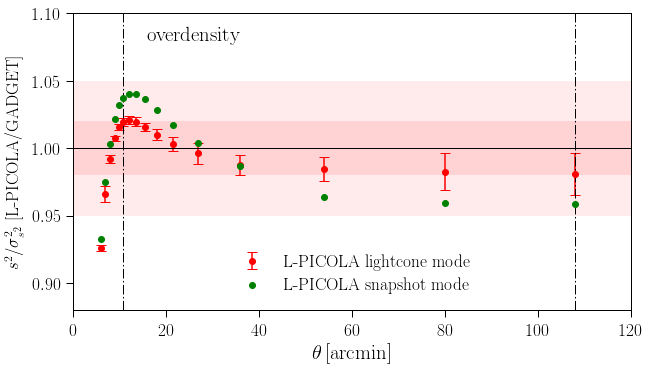}
\caption{Ratio $s^2 / \sigma_{s^2}^2$ for L-PICOLA maps over $s^2 / \sigma_{s^2}^2$ for GADGET-2 maps for the convergence (upper plot) and overdensity (lower plot). The results computed using the L-PICOLA lightcone mode represent the mean over 150 realizations together with their $1\sigma$-error. The dark-red area represent $2\%$ and the light-red area represents $5\%$ deviation from GADGET-2 and the vertical dot-dashed line marks a smoothing scale corresponding to $\ell = 10^2$ and $\ell = 10^3$.} \label{varvarplot}
\end{figure}

Next we compute moments as a function of smoothing scale $\theta$ to further quantify the differences between the maps. The variance is given by
\begin{equation}
s^2 = \frac{1}{N} \sum_{i=1}^{N} (q_i - \bar{q}) \, ,
\end{equation}
where $\bar{q}$ is the sample mean for elements $q_i = \kappa_i$ or $\delta_i $ within a sample of size $N$. Important for our analysis are the expectation value and the variance of this quantity. The expectation value is then given by $\left<s^2\right> =  (N-1) \mu_2/N$, where $\mu_2$ is the second moment of the distribution. We compute the variance of the variance estimator given by (Kenney \& Keeping \cite{Kenney})
\begin{equation}
\sigma_{s^2}^2 = \frac{(N-1)^2}{N^3} \mu_4 - \frac{(N-1)(N-3)}{N^3}\mu_2^2 \sim \frac{1}{N} (\mu_4 - \mu_2^2) \, ,
\label{varvar}
\end{equation}
where $\mu_2$ and $\mu_4$ are the second and fourth moment of the distribution respectively. Equation (\ref{varvar}) can be used to describe the non-Gaussianity of the distribution. In Figure \ref{varvarplot} we plot the relative difference of the quantity $s^2 / \sigma^{2}_{s^2}$ between maps computed using L-PICOLA and using GADGET-2 as a function of smoothing scale in arcmin. We find that $s^2 / \sigma^{2}_{s^2}$ for the convergence computed in the L-PICOLA lightcone and snapshot mode agrees within 2\% to GADGET-2 between smoothing scales corresponding to $\ell \sim 10^2$ and $\ell \sim 10^3$. The same quantity for the overdensity agrees within 2\% in the L-PICOLA lightcone mode and within 5\% in the L-PICOLA snapshot mode to GADGET-2.

\clearpage

\begin{figure}[htbp!]
\centering
\includegraphics[width=12cm]{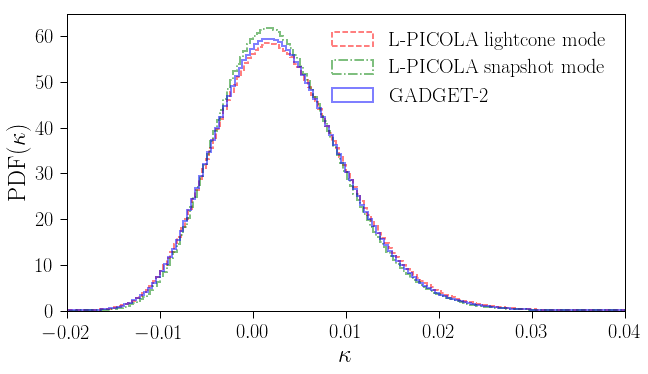}
\end{figure}

\begin{figure}[htbp!]
\centering
\includegraphics[width=12cm]{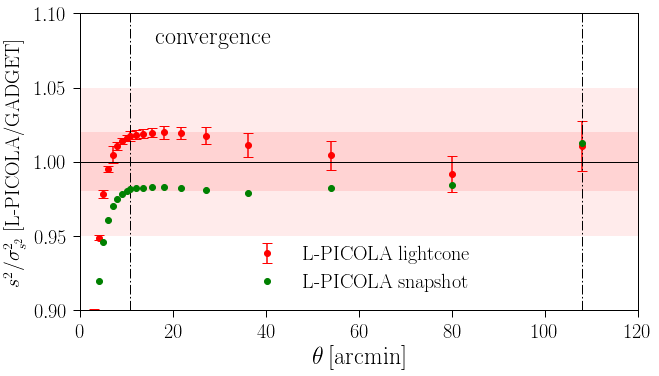}
\end{figure}

\newpage

\subsection{Cosmological Parameter Estimation}
\label{mcmc}

To further validate the accuracy of the covariance matrix we obtained, we constrain cosmological parameters with a Monte Carlo Markov Chain (MCMC) using  \textsc{CosmoHammer} (Akeret \textit{et al.} \cite{CosmoHammer}). We vary the two cosmological parameters $\Omega_m$ and $\sigma_8$ by fitting one realization of the convergence power spectrum $C_\ell^{\kappa \kappa}$ presented in section \ref{powerspectra} to a theoretical HALOFIT prediction for angular scales $10^2 < \ell < 10^3$ and using our estimated covariance matrix computed using 150 \textsc{UFalcon} maps based on L-PICOLA simulations in the lightcone mode. This test corresponds to an optimistic case of a full sky survey, without measurement noise and with only 2 free model parameters. It is performed to assess whether the accuracy of our covariance matrix is sufficient to derive cosmological parameter constraints. Furthermore, we modify our estimated covariance matrix by multiplying it with the angular scale-dependent ratio between the convergence power spectra based on L-PICOLA in the lightcone mode and GADGET-2. In this way, we obtain an estimate about how much our obtained constraints on $\Omega_m$ and $\sigma_8$ vary when the covariance matrix is rescaled in such a way by a few percent.

The analysis is done in the context of our fiducial cosmology with the parameters $h=0.7$, $\Omega_b = 0.045$ and $n_s = 0.961$ and with the likelihood (e.g. Nicola \textit{et al.} \cite{Andrina1})
\begin{equation}
\mathcal{L} (D | \theta) = \frac{1}{ [(2 \pi)^d \,\text{det} \, \text{cov} ]^{1/2} } e^{-\frac{1}{2} (C_\ell^{\text{sim}} - C_\ell^{\text{theory}})^{\text{T}} \text{cov}^{-1} (C_\ell^{\text{sim}} - C_\ell^{\text{theory}})} \, ,
\end{equation}
where $C_\ell^{\text{theory}}$ is the theoretical HALOFIT prediction for the spherical harmonic power spectrum computed within \textsc{PyCosmo} of dimension $d$ and $\text{cov}$ denotes the covariance matrix of the convergence power spectrum presented in section \ref{cov_matrix}. 

The convergence power spectra $C_\ell^{\kappa \kappa}$ computed in the L-PICOLA lightcone mode used for $C_\ell^{\text{sim}}$ and the covariance matrix have cosmic variance and shot noise contributions only and are computed from full-sky maps, such that effects of mask regions do not need to be taken into account. Note that since we vary only two cosmological parameters and do not add shape noise to the convergence power spectra entering the covariance matrix and to $C_\ell^{\text{sim}}$ and $C_\ell^{\text{theory}}$, this setup represents an optimistic case. 

We choose wide flat priors for the cosmological parameters $0.05 < \Omega_m  <  0.9$ and $0.2 < \sigma_8 < 1.6$. This is motivated by the choice of priors for the cosmological constraints analysis performed by the DES collaboration on the DES SV data \cite{Abbott}, which are close to those used in the DES Y1 analysis \cite{Y12}. Furthermore, we set the  the multiplicative bias parameter to $m=0$. The spherical harmonic power spectra vectors have been binned with $\Delta \ell = 24$.
 
\begin{figure}[htbp!]
\centering
\includegraphics[width=10cm]{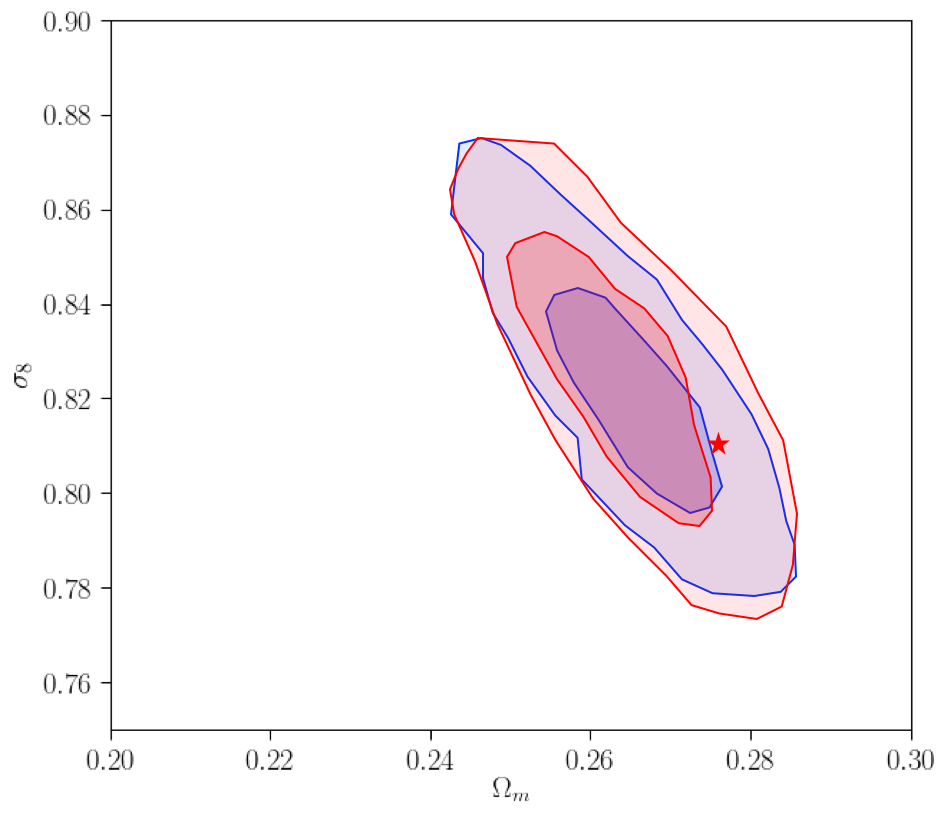}
\caption{Cosmological parameter constraints by varying $\Omega_m$ and $\sigma_8$ for weak lensing convergence for scales $10^2 < \ell < 10^3$ using a Monte Carlo Markov Chain. The two sets of contours depict the 68\% and 95\% confidence interval for the covariance matrix calculated from 150 convergence power spectra in the L-PICOLA lightcone mode (red) and for the covariance matrix modified by $C_{\ell}^{\mathrm{GADGET-2}} / C_{\ell}^{\mathrm{L-PICOLA}}$ (blue). The red star shows the fiducial values  $\Omega_m=0.276$ and $\sigma_8 = 0.811$.} \label{constraints}
\end{figure}

Moreover, we stress-test the obtained constraints by modifying the used covariance matrix through the ansatz:
\begin{equation}
\widetilde{\mathrm{cov}} ( \ell, \ell' ) \equiv f(\ell) f(\ell') \mathrm{cov} ( \ell, \ell' ) \, ,
\label{mod_cov}
\end{equation}
where $f(\ell) = C_{\ell}^{\mathrm{GADGET-2}} / C_{\ell}^{\mathrm{L-PICOLA}}$. In this way we modify the covariance matrix through a scale dependent function corresponding to the difference between the convergence power spectrum obtained using GADGET-2 and L-PICOLA in lightcone mode. In Figure \ref{constraints} we show the constraints on $\Omega_m$ and $\sigma_8$ by using the covariance matrix from section \ref{cov_matrix} and the modified covariance matrix given by equation (\ref{mod_cov}). The resulting constraints using the modified covariance matrix (blue) are only slightly smaller compared to the constraints obtained when using our originally estimated covariance matrix calculated from power spectra in the L-PICOLA lightcone mode (red). We conclude that such a rescaling of the covariance matrix does only minimally affect the constraints on $\Omega_m$ and $\sigma_8$, even for this optimistic case, and that the covariance matrix calculated using L-PICOLA is robust to changes on the percent level.
\\


\section{Conclusion}
\label{conclusion}

In this paper, we implemented a new framework to generate weak gravitational lensing maps in a fast, parallel way. Our pipeline is mainly optimized to estimate covariance matrices at the required resolution for scales between $10^2 < \ell < 10^3$ for wide field surveys such as the DES survey. The parallelized \textsc{UFalcon} code has been developed by using simulated matter density fields from the numerical code L-PICOLA, and then creates the all-sky past-lightcone by projecting the matter field at different redshifts onto the sphere. The code can also use GADGET-2 simulations for comparison. In order to optimize the mass resolution of the used simulation and thus the angular resolution of the created weak lensing maps, we adopted a nesting scheme to construct the lightcone.

In the present treatment, we calculated the convergence and galaxy overdensity maps using a single source redshift at $z_s = 1.5$. We compared the generated maps using L-PICOLA in different operating modes to the $N$-Body code GADGET-2 by considering the spherical harmonic power spectra, the PDF and higher-order statistics. The convergence power spectrum agrees with the nonlinear theory prediction within $5\%$ on angular scales $10^2 < \ell < 10^3$, whereas the overdensity power spectrum agrees within $5\%$ on $10^2 < \ell < 1600$. Compared to our results from using GADGET-2 simulated density fields, we find that the convergence and the overdensity power spectra from L-PICOLA agree within 5\% for $10^2 < \ell < 10^3$ to GADGET-2. Our results for the spherical harmonic power spectra are in good agreement with the findings from Izard \textit{et al.} \cite{COLA3} and capture nonlinear structures on slightly smaller angular scales than the $C_{\ell}^{\kappa \kappa}$ computed by Patton \textit{et al.} \cite{Patton}, although this is probably due to the higher value of {\tt NSIDE} used in our work.
Since our code is parallelized on all steps of the pipeline, an acceptable runtime of $\sim 3$ hours walltime is required to simulate the matter density field and construct the past-lightcone. This makes our framework suitable for generating a large number of weak lensing maps, as required to reduce the statistical noise in the covariance matrix. We computed the covariance of the spherical harmonic power spectrum based on 150 realizations using L-PICOLA. We observe off-diagonal clustering starting at $\ell \sim 2.5 \,\times \, 10^2$, indicating the onset non-Gaussian features of structure formation.

We extend the analysis of our constructed weak lensing maps by computing the PDF and the variance of the sample variance of the maps as a function of smoothing scale. We find a 2\% agreement between L-PICOLA and GADGET-2 for the variance of the sample variance for maps smoothed on scales $10^2 < \ell < 10^3$.
We derive an estimator for the covariance matrix computed using the full-sky spherical harmonic power spectra without shape noise. To test the obtained covariance matrix, we derive cosmological parameter constraints for a flat $\Lambda$CDM cosmology. We derive constraints for the parameters $\Omega_m$ and $\sigma_8$ from the convergence auto power spectrum and compare the results with the constraints obtained by applying a scale-dependent rescaling $C_{\ell}^{\mathrm{GADGET-2}} / C_{\ell}^{\mathrm{L-PICOLA}}$ of the used covariance matrix. The two obtained constraints do not show significant differences in our optimistic setup of only varying two cosmological parameters. We conclude that our \textsc{UFalcon} lightcone framework combined with the approximate code L-PICOLA is able to quickly produce sufficiently accurate convergence covariance matrices for angular scales in the range $10^2 < \ell < 10^3$ and can produce stable constraints in the $\Omega_m$-$\sigma_8$-plane for optimistic survey configurations.

Our past-lightcone framework thus opens a broad range of further possibilities. Foremost, it will be essential to include additional cosmological probes for the analysis of upcoming surveys. Moreover, it will be interesting to apply the pipeline presented here to models beyond $\Lambda$CDM.


\acknowledgments

We would like to thank Jorit Schmelzle, Janis Fluri, Tomasz Kacprzak, Risa Wechsler and Matthew Becker for very helpful discussions concerning the lightcone pipeline. Furthermore, we thank Uwe Schmitt for his help with the computing implementation. This research made use of \textsc{IPython}, \textsc{NumPy}, \textsc{SciPy}, \textsc{Matplotlib}, \textsc{Healpy}, \textsc{PyCosmo}, \textsc{CosmoHammer} and \textsc{Corner}. We acknowledge support by SNF grant 200021\_169130.

\end{document}